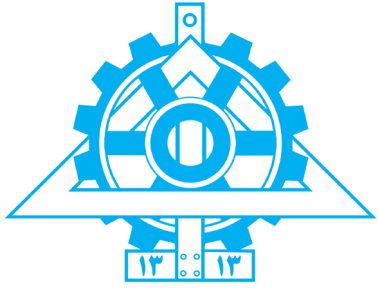
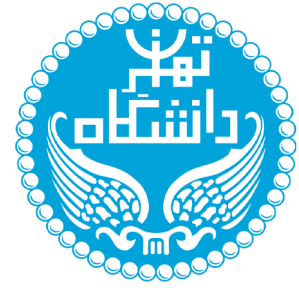

University of Tehran

College of Engineering

# An Agent-based Model Simulation Approach to Demonstrate Effects of Aging Population and Social Service Policies on Pensions Fund and Its Long-term Socio-economic Consequences

A thesis submitted to the Office of Academic Affairs

In partial fulfillment of the requirements for

the degree of Bachelor of Science in Industrial Engineering

**By:**

**Shayan Firouzian Haji**

**Supervisor:**

**Dr. Moeen Sammak Jalali**

**February 2024**

# Table of Contents

## Abstract

Agent-based modeling (ABM) has emerged as a powerful tool in social policy-making and socio-economics, offering a flexible and dynamic approach to understanding and simulating complex systems (Balmann, 2001). While traditional analytic methods may be less effective in unpredictable situations, ABM can provide valuable support for policy-making by generating large ensembles of scenarios and evaluating adaptive policies (Lempert, 2002). This approach has been applied in various fields, including economics, management, sociology, and politics, and has the potential to deepen our understanding of economic policy in the cooperative sector (Terano, 2007; Brodskiy, 2020).

# 1. Introduction

Agent-based modeling (ABM) is a highly effective tool for comprehending social and economic systems, as well as the emergence of patterns within them. By simulating the interactions of individual agents within a system, ABM enables the study of macroscopic patterns (Jensen, 2022) and thus has been applied to study socioeconomic systems. This flexible approach can capture complex behaviors and outcomes that arise from agents' interactions (Kwon et al., 2022), making it particularly helpful for analytically intractable models or situations where the payoff structure among agents can be calculated from their microscopic interactions (Di Russo et al., 2022). As a computational approach, ABM models economies as complex evolving systems with many interacting units over time (Richiardi, 2018) by simulating the actions and interactions of autonomous agents within a given environment. The approach aims to capture emergent phenomena that result from the interactions of individual agents, providing insights into complex system dynamics. Core concepts in ABM include representing respective agents, their decision-making processes, and their interactions with other agents and the environment.

Theories and models related to ABM often draw from complex systems theory, emphasizing social and economic systems' non-linear and dynamic nature. ABM also incorporates institutional economics, evolutionary theory, and social physics concepts to capture the adaptive and evolving nature of agents and their environments. The methodological framework of ABM involves defining agent behaviors, specifying the rules governing interactions, and simulating the system to observe emergent patterns.

## 1.1. Motivation & Significance

On May 2, 2023, Sajad Padam, the Director General of the Department of Social Insurance, provided an interview with a news agency in which he suggested that Iran may be compelled to sell Kish, Qeshm, and Khuzestan, similar to Greece, which sold about 100 islands, to fund pensioner salaries. This remark sparked social and political turbulence, leading to Padam's dismissal under public pressure two days later (*Controversial Statements about the Sale of Iranian Islands*, 2023).

Padam's reference to the sale of Greek islands underscores the economic and financial crisis that Greece encountered. In the 1990s, Greece experienced significant prosperity due to its access to the sea and the growth of shipping and tourism. The right to pension was attainable at the age of 55 with 30 years of work experience, and specific groups, such as civil servants, could retire at 52. However, with an aging population and an increase in the number of retirees, pension funds require government support to pay retirees, leading to heavy debt.

In 2008, the US financial crisis had a global impact on globalization, causing a recession in various countries, including Greece. This resulted in a decrease in people's purchasing power, a reduction in government revenues due to the decline in shipping and tourism, and Greek banks facing problems as most of the loans they provided to the Greek government were linked to pension funds. Consequently, the banks were at risk of defaulting on their loans. In an attempt to improve its financial situation, the Greek administration implemented austerity and expenditure-cutting programs, which resulted in a decline in public revenues and economic stagnation.

As a result of the large government debt and the inability to repay it, Greece sought international financial assistance in 2010. The EU and the International Monetary Fund (IMF) offered the Greek administration significant loans that required the implementation of austerity programs, including cutting government spending, reforming the pension system, privatizing the public sector, and raising taxes. These measures resulted in protests, strikes, and political unrest in Greek society, with some Greeks even migrating to other countries to avoid adverse economic conditions (*Story of the Greek Islands*, 2023).

Iran's economy experienced a significant rise in revenue in the 2000s due to various factors, including the previous administration's efforts to open up the hardline-controlled political environment of Iran on the global stage, laying the groundwork for global cooperation and international trade for the country. Furthermore, an increase in global oil prices resulted in an approximate income of $700 billion USD between 2005 and 2013 (*Economic Research and Policy Department Time Series Database*, n.d.).

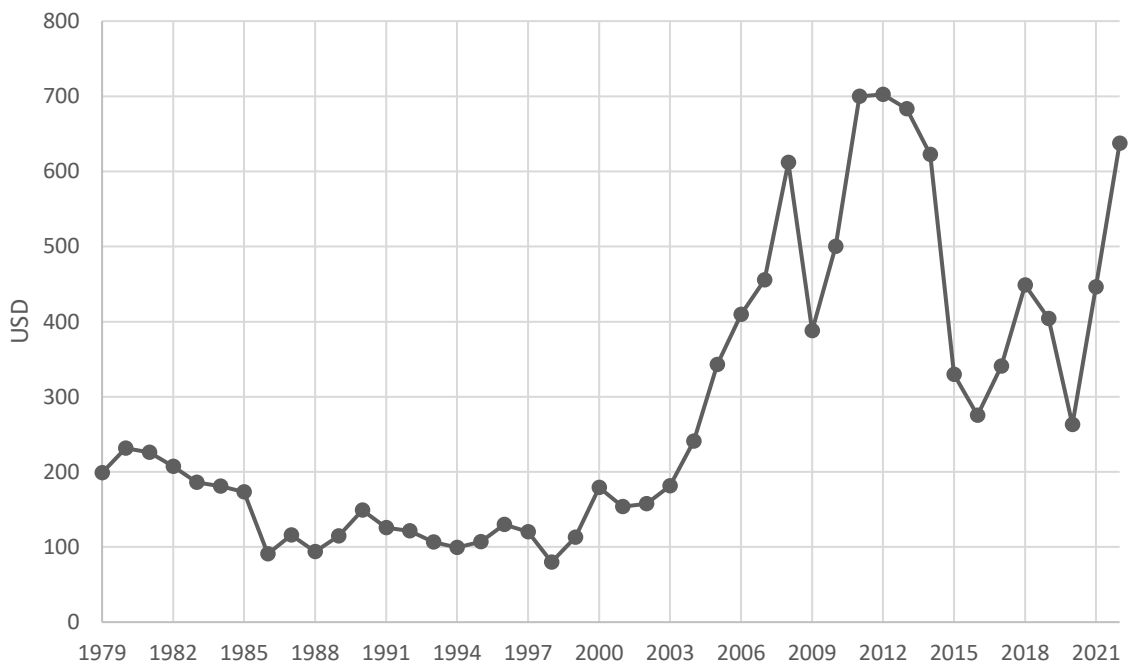

*Figure 1 Global price of crude oil for one cubic meter in current USD (since the Iranian Revolution until now)*

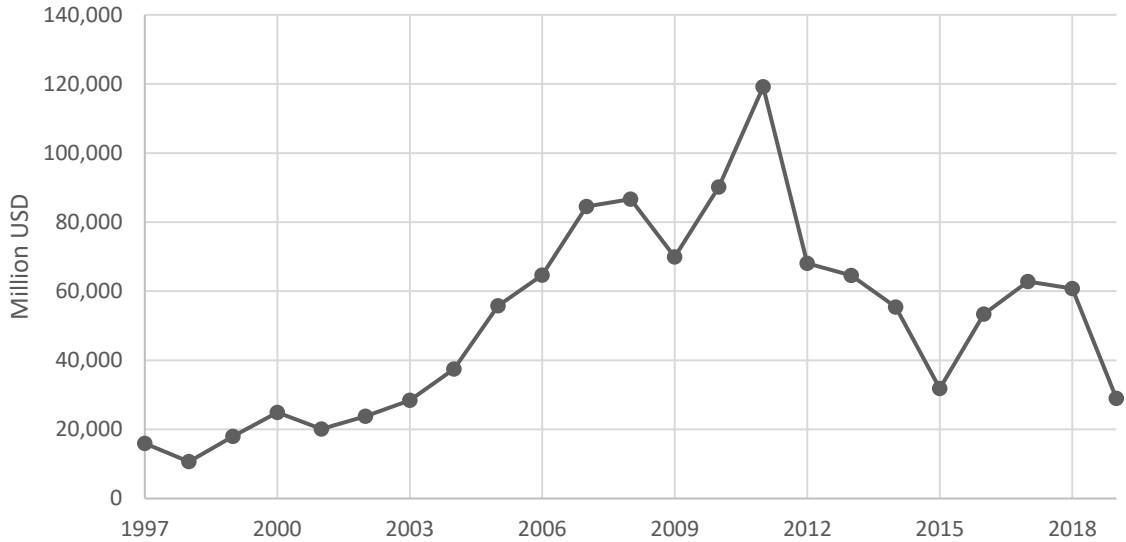

*Figure 2 Iran's revenue from oil export in million USD (from 1997 to 2019) (Crude Oil Prices, n.d.)*

Nonetheless, this was not a sustainable trend, and Iran's total income as a petro-economy was significantly impacted by individual and international sanctions. Expenditures on social welfare programs and unprofitable projects made it difficult for subsequent administrations to reverse these policies. Therefore, experts and officials have warned that observing ongoing patterns and trends within Iran's economic situation and demographic and social structure is critical. To the best of the authors' knowledge, no prior study incorporating all these aspects of the problem has been conducted. As the saying goes, "*Those who do not learn history are doomed to repeat it.*"; therefor, we conducted this research to examine the cause and evaluate its consequences.

**1.2. ABM in Socioeconomic Systems**

Interactions between individuals or entities within socioeconomic systems are governed by a set of rules, resulting in emergent phenomena at the macro level that are difficult to predict by examining individual components. ABM enables the representation of diverse agents, each with their own attributes and behaviors, and facilitates the simulation of their interactions over time. This can offer valuable insights into the dynamics of socioeconomic systems, including how local interactions can lead to global patterns, how changes in individual behavior can impact the overall system, and how the system evolves over time.

ABM is increasingly being utilized in a variety of fields, including economics, sociology, geography, and political science. It offers a powerful tool for understanding complex socioeconomic systems, making it an essential method for social scientists and policymakers. However, ABM models require careful design, calibration, and validation, and the accuracy of the results depends on the assumptions and data underlying the model. Therefore, it is necessary to employ ABM in conjunction with other methods and to interpret the results within the context of the model's limitations.

ABM is a bottom-up method of analysis where the global behavior of the system results from the local interactions of individual agents. Each agent in the model represents an autonomous entity, such as an individual, household, firm, or country, depending on the scale of the model.

The agents are typically heterogeneous, meaning they possess different attributes and behaviors that can change over time as a result of their interactions.

The environment in which agents interact can be represented in various ways, depending on the specific requirements of the model. It can be as simple as a grid where agents move around, or as complex as a realistic geographical landscape with various natural and artificial features. One of the key strengths of ABM is its ability to model complex adaptive systems, where the behavior of the whole is greater than the sum of its parts, and where interactions between parts lead to emergent properties at the global level. This makes ABM particularly useful for studying socioeconomic systems, which are inherently complex and adaptive. ABM is also more flexible than traditional mathematical models, allowing for more realistic representations of agent behavior and system dynamics, including non-linear interactions, path dependence, feedback loops, and adaptation.

However, despite these advantages, ABM presents certain challenges. Designing, calibrating, and validating ABM can be a complex and time-consuming process, often requiring large amounts of data. Results can be sensitive to specific assumptions and parameters used, and the complexity of models can make interpretation and communication of results difficult. While ABM is a powerful tool for studying socioeconomic systems, it should be used with caution and in conjunction with other methods. Like any tool, its effectiveness depends on the user's skill and knowledge. With careful application and interpretation, however, ABM can offer valuable insights into the behavior of complex socioeconomic systems and guide policy-making in these areas.

## 2. Background

### 2.1. Fundamentals

Agent-based modeling (ABM) has emerged as a significant methodological approach for analyzing complex adaptive systems (CAS) across various disciplines, including economics, sociology, and environmental science. The origins of ABM can be traced back to the work of Enrico Fermi in the 1930s, whose research on particle movement laid the foundation for simulating interactions between individual entities (Haldane & Turrell, 2019). Since then, ABM has developed into a powerful tool for studying systems in which the behavior of individual actors leads to emergent phenomena at the macro level, particularly in the fields of social sciences and economics (Sabzian et al., 2018).

A key feature of agent-based models is the heterogeneity of the agents. Each agent within the model can have distinct characteristics and behaviors, allowing researchers to simulate the diversity of real-world scenarios and the complexity of individual interactions. This heterogeneity is complemented by the autonomy of each agent. Agents function independently and make decisions based on their own internal rules and interactions with the environment and other agents (Brodskiy et al., 2021). Moreover, agents in ABMs exhibit adaptability, meaning they can learn and modify their behaviors over time in response to changes in their surroundings or actions by other agents (Brodskiy et al., 2021; Sabzian et al., 2018). Another notable aspect of ABM is the non-linearity of interactions among agents, which results in outcomes that are not directly proportional to the input behaviors. This non-linearity is critical to understanding the emergent properties of complex systems, where aggregate outcomes cannot be easily predicted from individual actions (Sabzian et al., 2018).

The mechanisms underlying agent-based models involve defining the agents, their behaviors, and the environment in which they operate. Agents are characterized by specific properties, such as wealth or preferences, and they engage in behaviors like communication or trading (Axtell, 2000; Brodskiy et al., 2021). The environment in which agents interact may be spatial or network-based, influencing how agents behave and the outcomes that emerge from their interactions (Brodskiy et al., 2021). The interactions between agents, governed by predefined rules, are crucial for producing emergent phenomena, where the collective dynamics cannot be easily deduced from the behavior of individual agents alone (Brodskiy et al., 2021; Sabzian et al., 2018).

The ability of ABM to illustrate emergent phenomena is one of its most valuable contributions to the study of complex systems. By modeling simple, local interactions between agents, ABM can demonstrate how these interactions give rise to complex global patterns, a key characteristic of CAS (Sabzian et al., 2018). Additionally, ABM provides a valuable framework for policy analysis. It allows researchers to simulate the effects of different policy interventions on economic systems, helping policymakers make more informed decisions by predicting the potential consequences of their actions (Haldane & Turrell, 2019; Kononovičius & Daniūnas, 2013). Furthermore, ABM offers behavioral insights by modeling individual behaviors and examining their influence on collective dynamics, such as market trends or social phenomena (Haldane & Turrell, 2019; Sabzian et al., 2018).

The methodology of ABM typically involves several key steps, starting with the design of the model. During this phase, researchers define the purpose of the model, the agents, their behaviors, and the environment in which they interact (Brodskiy et al., 2021). Following model design, the implementation phase uses programming languages or specialized software, such as NetLogo or Repast, to create the model (Abar et al., 2017). After implementation, the model is simulated to observe the outcomes of agent interactions and system-wide behaviors (Abar et al., 2017). Finally, the results are analyzed using statistical or qualitative methods to draw conclusions about the system's dynamics and validate the robustness of the findings (Axtell, 2000).

Agent-based models have been deployed in numerous fields, reflecting their versatility and practicality. In the social sciences, ABMs are commonly used to study phenomena such as cooperation, competition, and the formation of social networks (Brodskiy et al., 2021). In economics, ABMs play a crucial role in modeling market dynamics, consumer behavior, and the impact of economic policies (Haldane & Turrell, 2019; Kononovičius & Daniūnas, 2013). Environmental studies have also benefited from ABM, particularly in simulating ecological interactions and exploring the effects of human activities on ecosystems (Brodskiy et al., 2021).

The practicality of ABM is evident in its broad applicability across disciplines. ABM allows researchers to explore the intricacies of complex systems that traditional models often fail to adequately capture (Sabzian et al., 2018). Researchers can conduct experiments within ABM by manipulating variables and observing the outcomes, providing a controlled environment for hypothesis testing (Haldane & Turrell, 2019; Kononovičius & Daniūnas, 2013). Additionally, many ABM platforms feature visual interfaces that enhance the understanding of agent interactions and the system dynamics they produce (Abar et al., 2017).

In conclusion, agent-based modeling has significantly advanced the study of complex systems by offering a robust framework for understanding interactions and emergent phenomena. Its applicability across a range of disciplines underscores its value as a tool for both researchers and policymakers. As computational technologies continue to evolve, ABM is likely to become even more integral to studying and solving complex socio-economic and environmental challenges (Axtell, 2000). This literature review has highlighted ABM's unique capabilities, demonstrating its importance in contemporary research for modeling complexity and generating insights into dynamic systems.

## 2.2. Applications

ABM has become an increasingly prominent methodology across multiple fields due to its ability to simulate complex systems where individual behaviors lead to emergent phenomena. In economics, ABM has been extensively applied to model market dynamics, consumer behavior, and the impact of policies. By simulating heterogeneous agents interacting in markets, ABMs capture the complexities of real-world economic behavior. For example, these models can replicate business cycle fluctuations and income distributions, providing valuable insights into the processes generating such fluctuations (Dawid et al., 2018). Additionally, ABMs serve as effective tools for policy analysis. They allow researchers to explore the consequences of different economic policies by simulating their effects on resource allocation and market behavior, which may reveal unintended outcomes that traditional models cannot account for (Axtell, 2000; Brodskiy et al., 2021). Simulating the impacts of fiscal policies can offer policymakers a better understanding of the dynamic responses of economic agents, enabling them to evaluate potential scenarios more comprehensively (Dixon, 2011).

In the social sciences, ABM is employed to study social interactions, cooperation, and conflict resolution, allowing for an in-depth understanding of how individual behaviors contribute to collective social phenomena. This is particularly evident in studies of segregation or cooperation, where ABMs can demonstrate how individual choices lead to large-scale patterns in society (Sabzian et al., 2018; Silverman et al., 2011). One classic example is Schelling's model of residential segregation, which shows how seemingly innocuous individual preferences can culminate in substantial social segregation (Silverman et al., 2011). Additionally, ABMs are well-suited for modeling complex adaptive systems, as they allow agents to be represented as adaptive entities. Through these models, researchers can explore how social norms and behaviors evolve over time, capturing the dynamic nature of social systems (Sabzian et al., 2018).

ABM has also found widespread use in environmental studies, particularly in modeling ecological interactions and land-use changes. ABMs can simulate the interactions between various agents, such as species or humans, and their environment, offering insights into ecosystem dynamics and resource management (Parker et al., 2003). For example, these models can demonstrate how human activities affect biodiversity and ecosystem services, allowing researchers to predict long-term ecological impacts (Parker et al., 2003). In studies of land-use and land-cover change, ABMs are valuable for simulating how the decisions of individual land managers contribute to broader environmental outcomes. Such simulations can inform better planning and policymaking by providing a detailed view of how localized actions scale up to affect regional or global environmental systems (Parker et al., 2003).

In the field of public administration, ABMs enhance both theory development and policy evaluation by simulating complex governance structures and collaborative decision-making processes. These models enable researchers to test hypotheses about public sector dynamics, such as how different governance arrangements influence policy outcomes and citizen engagement (Choi & Park, 2021). By modeling the interactions among various stakeholders, ABMs provide insights into the effectiveness of different governance strategies and their implications for public policy (Choi & Park, 2021). Such simulations are particularly useful for evaluating how collaborative governance frameworks affect policy implementation and the responsiveness of public institutions to citizen needs.

In healthcare, ABMs are applied to model disease spread and the dynamics of healthcare systems. In epidemiology, ABMs simulate the spread of infectious diseases, incorporating individual behaviors and interactions to better understand outbreak dynamics. These models are crucial for informing public health interventions by predicting how diseases spread across populations and the likely impact of different containment strategies (Abar et al., 2017; Brodskiy et al., 2021). ABMs are also used in healthcare resource management, where they simulate patient flow and resource allocation within healthcare systems. This enables healthcare administrators to optimize service delivery and improve patient outcomes through more effective resource planning and allocation (Abar et al., 2017; Brodskiy et al., 2021).

In management science, ABM is utilized to study organizational behavior and decision-making processes within firms. These models can simulate interactions within organizations, allowing researchers to explore how individual behaviors influence overall organizational performance (Sabzian et al., 2018). By modeling scenarios such as team dynamics, leadership effects, and the impact of organizational structures on performance, ABMs provide valuable insights into the internal dynamics of organizations. Furthermore, ABMs are useful for examining strategic decision-making in competitive environments. By simulating how firms adapt their strategies in response to market changes and competitor actions, ABMs contribute to a deeper understanding of strategic interactions and market competition (Sabzian et al., 2018).

In conclusion, agent-based modeling serves as a versatile and powerful tool across various disciplines, providing valuable insights into complex systems characterized by individual interactions. ABM contributes to enhancing theoretical understanding, as well as offering practical applications in economics, social sciences, environmental studies, public administration, healthcare, and management science. As computational capabilities continue to advance, the potential applications of ABM will expand further, solidifying its role as an essential methodology for both researchers and practitioners.

**2.3. In Socio-economic Systems**

ABM has emerged as a versatile tool for studying complex systems across various domains, including social sciences, demographics, economics, finance, and public policy. By simulating the interactions of individual agents and observing the resulting emergent phenomena, ABM allows researchers to model dynamic and heterogeneous behaviors that are often difficult to capture using traditional approaches. This review highlights the key applications of ABM in these fields, emphasizing its advantages and suitability for each.

In the social sciences, ABM is widely used to understand social dynamics by simulating interactions such as cooperation, competition, and the evolution of social norms. Schelling's model of residential segregation is a classic example where individual preferences lead to large-scale social outcomes (Axtell, 2000). ABM is particularly valuable for studying complex adaptive systems, where individual actions result in emergent social behaviors (Axtell, 2000). By modeling the interactions between diverse agents, ABM can explore how social norms evolve and provide insights into the dynamic and heterogeneous nature of social interactions. Its ability to represent diverse agents and their behaviors offers a distinct advantage over traditional models, which may not capture the nuanced and emergent phenomena seen in real-world social systems (Axtell, 2000).

In demographics, ABM is applied to study population dynamics, such as fertility rates, marriage patterns, and migration. This approach is also used for micro-simulations, which model individual life histories to gain detailed insights into population behavior (Bijak et al., 2013). ABMs are flexible and can incorporate various demographic factors, making them highly suitable for policy analysis and scenario testing. Another significant advantage is that ABMs reduce dependence on large datasets, allowing researchers to explore demographic trends and behaviors without relying on extensive data collection, which is often costly and time-consuming (Bijak et al., 2013).

ABM has made substantial contributions to economics by modeling market dynamics and enabling policy analysis. The ability of ABM to simulate trading behaviors and market fluctuations offers a more nuanced understanding of economic systems than traditional models (Brodskiy et al., 2021). ABM excels at capturing non-linear relationships among agents, a common feature of real-world economic systems (Brodskiy et al., 2021). Through the simulation of agent interactions, ABM can illustrate how individual behaviors aggregate to create macroeconomic outcomes such as business cycles, market inefficiencies, and income distribution. This feature is particularly valuable for studying emergent phenomena that may not be apparent when using more simplified or aggregate economic models (Brodskiy et al., 2021).

In finance, ABM is used to model market behavior, including the formation of bubbles, crashes, and volatility patterns (Haldane & Turrell, 2019). These models are instrumental in studying risk assessment by simulating the actions of various market participants under different conditions. ABMs provide a framework for exploring how heterogeneous financial agents, each with distinct strategies and levels of information, influence market dynamics (Haldane & Turrell, 2019). Moreover, ABMs allow researchers to investigate how agents dynamically adapt their strategies based on changing market conditions, offering insights into the complex nature of financial systems and risk management practices (Haldane & Turrell, 2019).

ABM has also proven useful in public policy and administration, where it is employed to simulate the effects of policy decisions across various sectors, such as healthcare, energy, and urban planning. ABM models the interactions among stakeholders and helps researchers analyze how collaborative governance processes unfold (Chappin et al., 2012). The strength of ABM lies in its ability to represent the intricate and interconnected nature of policy environments, where multiple actors with differing objectives and strategies interact (Chappin et al., 2012). This capability enables policymakers to simulate various scenarios, assess the potential impacts of policy options, and improve decision-making by testing and comparing the outcomes of different interventions (Chappin et al., 2012).

In conclusion, agent-based modeling is a powerful and adaptable methodology with broad applications in the social sciences, demographics, economics, finance, and public policy. Its ability to represent heterogeneous agents, simulate dynamic interactions, and capture complex system behaviors makes it an essential tool for studying a wide range of phenomena. As computational resources continue to improve, ABM's applicability and effectiveness are likely to expand, making it an indispensable tool for researchers and decision-makers alike.

## 3. Methodology

### 3.1. Problem Statement

It has come to our attention that a significant number of Iran's pension funds, precisely 15 out of 17 (*Bankruptcy of Pension Funds in Iran*, 2023), have become insolvent, leading to continuous governmental bailouts and causing a perennial source of friction between the government and the elderly. Iran's allocation for pensions has risen from 12 to 15 percent of the national budget over the past decade ("The Scale of the Crisis in Pension Funds," 2023), yet this hides a profound crisis that reflects three adverse trends. The first trend results from decades of deficient growth and double-digit inflation that have left real per capita income to decrease drastically. As a result, many retirees find themselves without personal savings and solely dependent on their monthly pensions, whose dollar value has fallen significantly. The minimum monthly pension payment has decreased by roughly 25 percent from where it stood ten years ago, at about $160, to the current $120 (Ghasseminejad, 2023).

Secondly, the aging of Iran's population has led to an increase in the number of retirees and pensioners, and they are living longer, which makes them require pensions for more extended periods than the country's retirement system was designed to handle (Doshmangir et al., 2023). Meanwhile, the active workforce that pays into retirement funds is estimated to be around 23 million due to decreasing birth, employment, and labor participation rates, which compounds the issue further (Ghasseminejad, 2023). High youth unemployment exacerbates the problem, further straining available resources.

Given the current dire state of the pension funds in Iran, pension funds are an integral component of a nation's social security system, providing financial support to retired citizens. However, these funds face increasing pressure worldwide due to shifting societal, demographic, and economic patterns. In Iran, these pressures are particularly severe, and the National Pension Fund, the country's second-largest pension fund, covering almost 880,000 insured workers, is facing a significant risk of bankruptcy due to the imbalance between the number of pensioners and contributors (Stone, 2023). Additionally, corruption and mismanagement have further exacerbated the situation, making it even more challenging to sustain the pension fund. The National Pension Fund has been utilized as a source of cheap loans for government projects, which has resulted in a shortage of funds. Moreover, the need for more transparency in the management of pension funds has made it challenging to monitor their financial health and identify potential issues. The imbalance between the number of pensioners and contributors is expected to grow, further exacerbating the problem.

The potential for a pension crisis and escalating national debt in Iran is not just a theoretical concern but a looming reality with severe long-term consequences. Consequently, it is imperative to study the specific interplay of these factors within the Iranian context and develop effective strategies to prevent such outcomes. Such research is both feasible and necessary, given the availability of demographic and economic data and the critical need to ensure the sustainability of pension funds. Addressing this issue is crucial in contributing to the stability and prosperity of Iran's economy and the well-being of its aging population. The urgency of this issue is underscored by the recent protests by Iranian retirees, who have been demonstrating against the ongoing erosion of their pensions and the crisis of pension funds. These protests highlight the real-world impact of the pension fund crisis and the urgent need for research and solutions. Thereby, it is clearly shown that it is necessary to explore viable solutions that address the systemic issues underlying the pension crisis and mitigate its impact on the economy and its citizens.

By studying the patterns of society, demography, and economy contributing to Iran's pension crisis and national debt, this research aims to provide valuable insights and potential solutions to this pressing issue. The findings of this research could inform policy decisions and strategies to mitigate the pension crisis, thereby contributing to the financial stability and social harmony of the nation.

## 3.2. Background

The particular issue of pensions, commonly known as the pensions crisis, has been subjected to several studies. However, most have either conducted case-wise studies by referring to one specific case of such crisis in the local context and within its age, or they have provided theoretical or practical frameworks to tackle the technical difficulties of these crises, such as managing financing funds, reforming regulations, and collaboration and transparency. Some proposed making changes to contributions and benefits, which is straightforward and can have immediate effects on the pension fund's financial stability. However, it can be politically challenging to implement, especially if it involves reducing benefits. It may also disproportionately affect certain groups, such as lower-income workers. Another approach to enhancing the performance of pension funds is to address governance shortcomings to increase public trust. It should be noted that this approach requires significant political will and may face resistance from vested interests. Adding some flexibility to pension rights may also seem a viable option, with the potential to help safeguard members' pension rights, particularly in defined-contribution schemes, but it may be difficult to implement in practice, and there may be legal or regulatory barriers to making pension rights more flexible. In the same context, it is necessary to calibrate assumptions with the actual facts. Revising actuarial assumptions can help ensure that pension calculations reflect current demographic and economic realities. Still, making changes toward actuarial assumptions can be controversial and may face resistance from stakeholders. Some researchers argue that diversifying investments and sources of income can help protect pension funds from market volatility, but this strategy has the disadvantage of being complex to implement and hardening shares to manage. Collaboration among stakeholders can also help ensure that all stakeholders' interests are considered and that solutions are sustainable in the long term. However, It can be challenging to get all stakeholders to agree on a common approach, and the process can be time-consuming.

These approaches are not mutually exclusive and can be combined in various ways to address the specific challenges faced by different pension systems. But what they all lack is a fundamental explanation of the nature of the crisis as a result of collective behavior and presenting a comprehensive picture of the evolution of it as a policy-made phenomenon. This is what this study refers to as emergence and aims to analyze its mechanisms as a consequence of collective behavior.

### 3.3. Agent-Based Modelling Approach

ABM offers several distinct advantages over conventional methods (like what is previously mentioned) when studying social sciences and behavioral economic phenomena. As our approach, ABM allows researchers to model complex systems with a high degree of heterogeneity and interaction among individual agents. Unlike traditional aggregate-level models, ABM captures the emergent properties that arise from the interactions and behaviors of diverse agents, providing a more realistic representation of real-world dynamics. ABM also facilitates the exploration of non-linear relationships and feedback loops within social systems. By incorporating individual-level decision-making processes and adaptive behaviors, ABM can simulate the dynamic evolution of social phenomena over time. This capability enables researchers to examine the emergence of unexpected outcomes and tipping points, which are often overlooked by linear models. Furthermore, ABM provides a flexible framework for incorporating empirical data and theoretical insights from various disciplines. Researchers can calibrate their models based on empirical evidence and validate them against real-world observations, enhancing the credibility and applicability of their findings. Additionally, ABM allows for the integration of insights from psychology, sociology, and other behavioral sciences, enabling a more comprehensive understanding of human behavior in complex social contexts.

In the context of studying the emergence and evolution of pension crises, ABM offers several advantages over conventional approaches. Pension systems involve a multitude of interacting factors, including demographic trends, economic conditions, institutional arrangements, and individual behaviors. ABM allows researchers to model these factors at the micro-level, capturing the diverse range of decision-making processes and interactions that shape the dynamics of pension systems. By simulating the behavior of individual agents, ABM can elucidate how changes in demographic patterns, retirement decisions, investment strategies, and policy interventions influence the sustainability and resilience of pension systems.

Moreover, ABM enables researchers to explore the potential consequences of different policy measures and reform options, providing valuable insights for policymakers and stakeholders.

## 4. Modelling Process

The base of this model was first conceptualized and mathematically developed by Epstein & Axtell (1996) and later implemented in NetLogo software by Li & Wilensky (2009). The base model was solely concerned with economic inequality in society and its origins and evaluating Pareto's theory about the distribution of wealth (Merritt, 1898). Although the base model is too primitive, it opened a new door to the field of complex systems analysis and emergent phenomena. Thus, we built our model upon the basics of what they have developed and developed it to best suit our objectives, which are mainly around assess the status of the pension fund.

**4.1. Elements**
**a) Environment**
Our *environment* is simply a spatial distribution of a generalized resource that, from now on, we call *sugar*, which agents must eat to survive. The *environment* is a two-dimensional coordinate lattice. At every point $(x, y)$ on the lattice, there is both a *sugar level* and a *sugar capacity*, the capacity being the maximum value the sugar level can take at that point. Some points might have no sugar (a level of zero) and low capacity, and others might have no sugar but large capacity- as when agents have just harvested all the sugar- while other sites might be rich in sugar and near capacity. The *environment* is arranged on a $50 \times 50$ lattice with the *sugar level* at every site initially at its total capacity value.

The initial *sugar level* is highest at the peaks in the northeast and southwest quadrants of the grid- where the color is most yellow- and falls off in a series of terraces. The *sugar level* ranges from a maximum of 4 at the peaks to zero at the extreme periphery.

The *environment* wraps around from right to left (that is, were you to walk off the screen to the right, you would reappear at the left) and from top to bottom, forming a doughnut-like shape we call a torus. In our model, autonomous *agent*s inhabit this *environment* and constantly

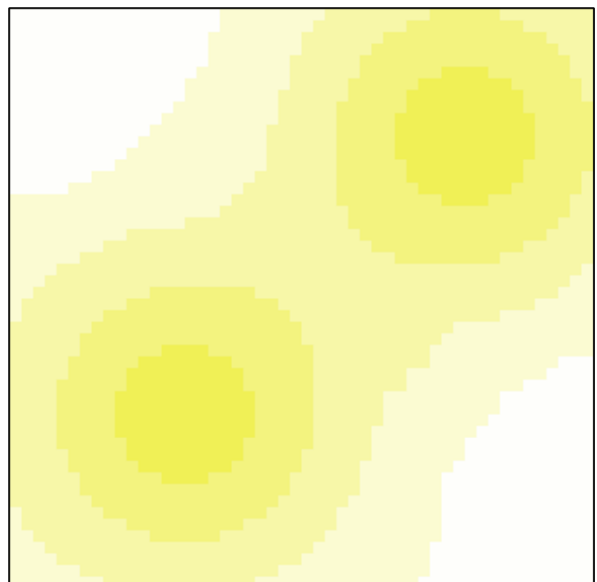
*Figure 3 Landscape of our enviroment*

collect and consume *sugar*. As a rule for how the sugar regenerates- how it grows back after the agents harvest it -we set it in a way that grows back at a rate of one unit per time step.

**b) Agents**

Just as there is an initial distribution of *sugar*, there is also an initial population of *agent*s. We give these *agent*s the ability to move around the *environment* to perform various tasks. They move, gather *sugar,* and eat it. These actions require that each *agent* have internal *state*s and behavioral *rule*s. Therefore, we describe these as follows.

**b.1) Agents' State**

Each *agent* is characterized by a set of fixed and variable *state*s. For a particular agent, its genetic characteristics are fixed for life, while its wealth, for instance, will vary over time. One state of each agent is its location on the landscape. At every time, each *agent* has a position given by an ordered pair (x, y) of horizontal and vertical lattice coordinates, respectively. No two *agent*s are allowed to occupy the same position. As *agent*s are initially distributed randomly all around the *environment*, some *agent*s are born high on the peaks of the sugar mountains.

Others start out in the sugar badlands, where sugar capacities are deficient. One might think of an *agent*'s initial position as its environmental endowment.

Each agent has a genetic endowment consisting of a sugar *metabolism* and a level of *vision*. *Agent*s have different values for these genetic attributes; thus, the agent population is heterogeneous. The *agent*'s *metabolism* is simply the amount of sugar it bums per time step or iteration. *Metabolism*s are randomly distributed across agents with a minimum of 1 and a maximum of 4. *Agent vision* is also randomly distributed. *Agent*s with *vision v* can see *v* units in the four principal lattice directions: north, south, east, and west. *Agent*s have no diagonal *vision*. This lack of diagonal *vision* is a form of imperfect information and represents bounded rationality. In our model, random values for *vision* ranging from 1 to 6 are initially assigned to *agent*s. All first-generation *agent*s are randomly given some initial endowment of sugar between 5 and 25, which they carry with them as they move around the *environment*. *Sugar* collected but not eaten -what an *agent* gathers beyond Its *metabolism*- is added to the *agent*'s *sugar* holdings. There is no limit to how much *sugar* an individual *agent* may accumulate. As for their life expectancy, each *agent* is randomly set to have a **Maximum Age** (*INT*) between 60 and 100, after which it dies and gets removed from the *environment*.

**b.2) Agents' Rules**

The *agent*s are also given multiple rules to behave accordingly, like movement *rule*s. Movement *rule*s process local information about the landscape and return rank orderings of the patches by the amount of *sugar* present at each patch within an *agent*'s *vision*. This can be expressed as the following instruction, which is a gradient search algorithm:

1. Look out as far as *vision* permits in the four principal lattice directions and identify the unoccupied patch(es) having the most *sugar*;
2. If the greatest *sugar* value appears on multiple patches, then select the nearest one;
3. Move to this patch;
4. Collect all the *sugar* at this new position.

Succinctly, this *rule* amounts to this: From all lattice positions within one's *vision*, find the nearest unoccupied position of maximum *sugar*, go there, and collect the *sugar*. At this point, the *agent*'s accumulated *sugar* wealth is incremented by the *sugar* collected and decremented by the *agent*'s *metabolism* rate. Suppose at any time the agent's sugar wealth falls to zero or below. In that case, that is, it has been unable to accumulate enough *sugar* to satisfy its metabolic demands- then we say that the *agent* has starved to death, and it is removed from the *environment*. Each agent is permitted to move once during each time step. The order in which *agent*s move is randomized at each step.

Additionally, the **Number of Children** (*INT*) attribute of each agent determines how many children each agent should have in the course of their life, specifically when they reach the **Age to Reproduce** (*INT*), in which they will have breed as many children as they were determined to have simultaneously. First generation excluded (they are all red), agents who are under breeding age and have no children yet are colored in green, and as soon as breeding, they will turn red. As a simplified form of inheritance, each born *agent* would have the same *vision* and *metabolism* as its parent, and its initial *sugar* level would be an equal share of the parent's wealth; the parent would pass half of their *sugar* savings to be equally divided among their children. Therefore, the initial *sugar* level of a newborn *agent* can be derived according to the formula below:

$$Sugar_{child} = \frac{Sugar_{parent}}{2 \times (N_{child\ siblings} + 1)} = \frac{Sugar_{parent}}{2 \times N_{children}}$$

In the simulation, agents retire at a certain age specified by **Retirement Age** (*INT*), and they will be hidden from within the environment because they stop moving around and collecting sugar, so working agents can move to the positions where retired agents stay and collect the sugar. Therefore, every agent has a state variable indicating its **Retirement Status** (*BIN*).

The sugar retired agents consume based on their metabolism is taken from the pension fund, and they consume sugar from their savings if the fund is empty.

Before being retired, agents are taxed an amount of the sugar retrieved in each timestep that goes to the pension fund. This is represented by the variable **Pension Tax** (*INT*) in the simulation as well as a **Fixed Fee** (*FLOAT*) of their income.

Suppose the **Social Services** (*BIN*) parameter is activated. In that case, working agents can get sugar from the fund to avoid dying as a consequence of running out of sugar, but, unlike retired agents, they still move and get sugar, considering their productivity if productivity decay is activated.

Because all of the initial population has an initial age of 0, there would be a first massive wave of retirements when reaching the retirement age.

The **Productivity Decay** (*BIN*) parameter allows agents to recollect only part of the sugar in a patch based on a productivity-decay rate that depends on age. These values have been obtained using a Univariate Akima Interpolation over productivity values per age given according to Bertoni et al. (2015).

### 4.2. Scenarios

With all the ingredients in hand, we can now define different scenarios to run our model under different circumstances. Many variables can be subject to further investigation. However, for the purpose of this study, the authors are interested in a few of them, namely the status of **Social Services** and **Productivity Decay** and the kind of probability distributions of random variables.

Productivity decay and social services can be either on or off, and the distribution of random values such as *vision*, *metabolism*, and *Age to Reproduce* can follow either a uniform distribution within the range of a lower and upper bound ($U(l, u)$) or a normal distribution with a mean and a standard deviation ($N(\mu, \sigma)$). For simplicity, we define a notation to express each scenario in a short form as **Social Services** and **Productivity Decay** are either *ON* or *OFF*, and the distribution of those three random values are either *N* or *U*. For example, a scenario in which there is social services available but productivity decay is not taken to account, and vision, metabolism, and Age to Reproduce are uniformly distributed can be shown as $S(ON, OFF, U)$.

It is worth noting that in the case of a normal distribution, parameters are as follows: $vision \sim N(3.5, 0.8)$, $metabolism \sim N(2.5, 0.5)$, and $Age\ to\ Reproduce \sim N(32.5, 5.8)$.

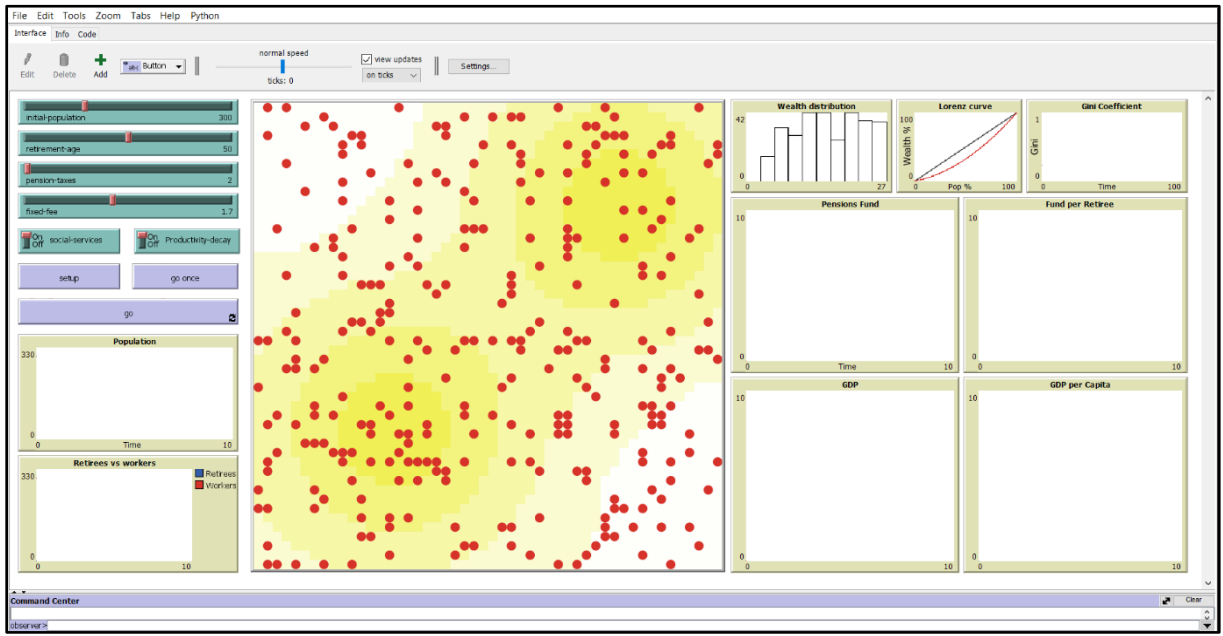

*Figure 5 Model is designed an sat up, ready to run. NetLogo software version 6.3*

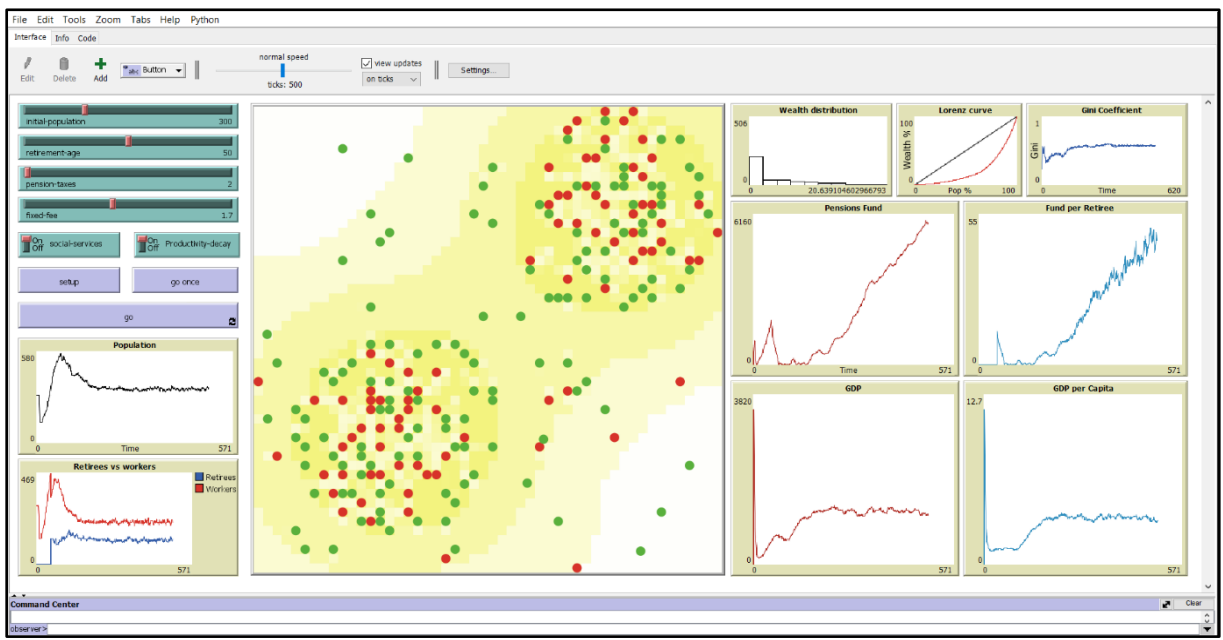

*Figure 4 Model is running, currently at 500th tick (timestep)*

### 4.3. Run Simulations

In order to narrow down the results we want to reach and also be able to perform sensitivity analysis on the results, four variables in our model are of special interest regarding the scope of the research: Maximum Age, Retirement Age, Pension Tax, and Fixed Fee. We will analyze these four variables in two sense-making pairs for the sake of visualisability, once as *Maximum Age vs. Retirement Age* and another as *Pension Tax vs. Fixed Fee*.

We also enriched these simulations with indicators of macroeconomics such as *GDP* and *GDP per Capita*, as well as *populations* in detail and indicators of economic inequality like the *Lorenz Curve* and the deriving *Gini Coeficcient*, and also a *histogram* to show the distribution of wealth in society in ten deciles.

## 5. Results and Discussion

Under $S(0,0,U)$, we ran the model 5 times for each pair value of *Pension Tax* and *Fixed Fee*, each time for 500 artificial timesteps (ticks) to extract the macro-trends, patterns, and potential tendencies. Results are shown in Fig. 7.

Knowing that the genetic characteristics in populations usually follow some kind of normal distribution, for the second set of simulations we configured $vision \sim N(3.5,0.8)$, $metabolism \sim N(2.5,0.5)$, and $Age\ to\ Reproduce \sim N(32.5,5.8)$ in order to gain more realistic results. The *mean* and *standard deviation* for each random variable are calculated so that the range of $\mu \pm 3\sigma$ for each normally distributed variable overlaps with the range of its uniform distribution. As was expected for this more realistic case, indicators of interest, especially *Fund per Retiree* and *Population*, show some cyclic behavior. It aligns with the theory of secular cycles, which is presented by some contemporary academicians (Turchin & Nefedov, 2009) and their intellectual forefathers such as Malthus and Ricardo (Abel, 1980). The model offers a systematic approach to explore the impact of genetic makeup on carrying capacity. By altering the genetic composition of the agent population, the model generates various distributions of vision and metabolism, resulting in stochastic fluctuations in population levels. Through this analysis, we gain a deeper understanding of the model's resilience and susceptibility to changes in agent traits.

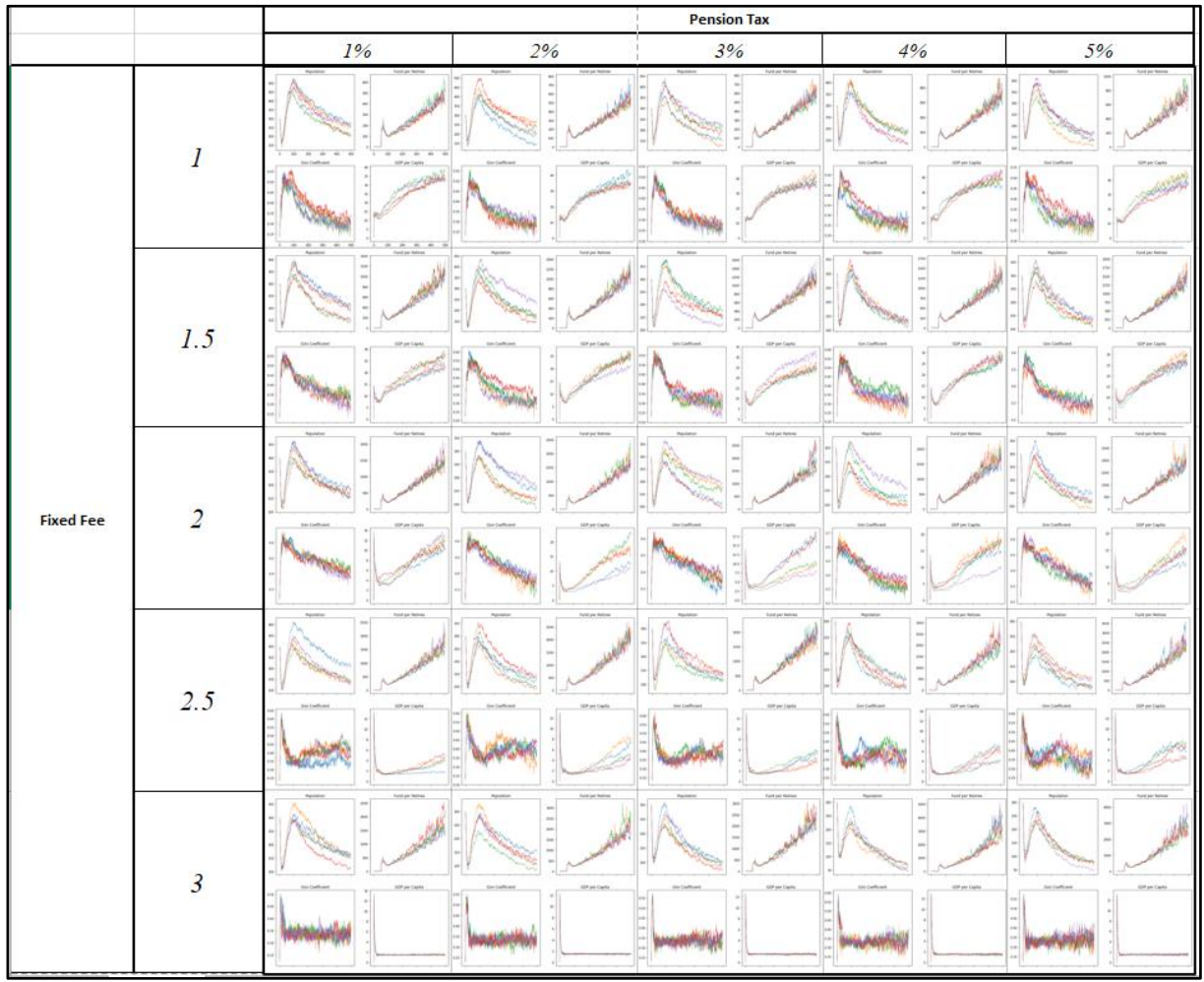

*Figure 7 Population, Gini Coefficient, Fund per Retiree, and GDP per Capita under S(0,0,U) with different values for Pension Tax and Fixed Fee*

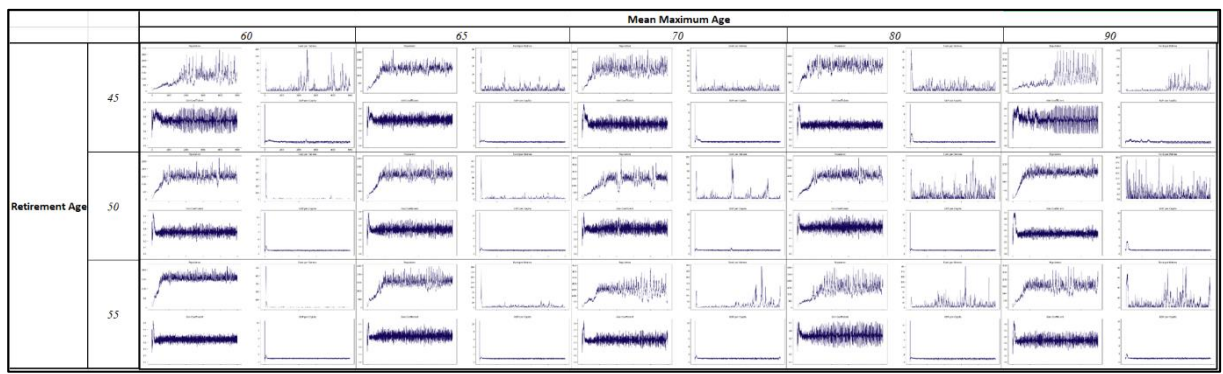

*Figure 6 Population, Gini Coefficient, Fund per Retiree, and GDP per Capita under S(1,1,N) with different values for Retairement Age and mean Maximum Age*

The notion of physical carrying capacity pertains to the design aspect and entails determining the maximum number of use units accommodated in a specific area (McLachlan & Brown, 2006). Therefore, carrying capacity is a central concept in our study. In other words, it refers to the capacity of a habitat or environment to sustain use units' existence, like human life, by providing necessary resources (Geores, 2001). By following the specified laws of genetic inheritance, the model showcases a progression in the overall vision and metabolism of the population over multiple generations. This advancement towards increased fitness, characterized by heightened vision and lowered metabolism, is a clear example of natural selection at work. As the population nears its maximum capacity, there is a notable shift towards these beneficial traits, signifying a gradual enhancement in the fitness of the population with each passing generation.

Because all of the initial population has an initial age of zero, there would be a massive spike in all indicators at the early stages of each running session, representing the flux of the first generation. It is further reinforced by the fixed retirement age, which is set to be the same constant value for the whole population. Despite the chaos in the early stages, the model eventually stabilizes as less successful agents are eliminated, leaving only those who occupy the best positions. The agent-based model simulates how agents behave within a terraced landscape, where they tend to stick to certain ridges that represent optimal positions and, over time, reach a steady-state configuration. This phenomenon reveals the natural selection process within the model, resulting in the emergence of stable states.

Finally, our model reveals a decentralized harvesting rule, where agents independently determine their local harvesting strategies based on their immediate surroundings. This self-organizing behavior results in macroscopic patterns of resource utilization, ultimately leading to an increase in GDP, GDP per capita, and improvements in social welfare. Our model presents practical evidence showcasing how collective self-interest maximization can enhance the overall situation and promote the greater good (Hayek, 1960). As a result, it can be a valuable contribution to debates on actual social decision-making and legislation (Phelps, 2009).

As a follow-up for further research building upon the results of this work or have a similar approach in methodology, one can consider deploying the following considerations:

- Taking into account the inflation rate on salaries, fees, and prices, as well as the deposit of public funds.
- Assigning different retirement ages to different agents in a single run.
- Determining a non-zero age for the initial population.

## 6. Conclusion

The Gini coefficient was not significantly affected under any scenario and combinations of different conditions, which means that economic inequality was not different under different circumstances, at least for the remaining population. It is demonstrated that in the absence of social services and hypothetical conservation of productivity level in individuals, the carrying capacity (population) would quickly drop for the increasing amount of fixed fees and also pension taxes, although at a much slower pace for the latter. The same is also true for GDP per capita. The depopulation is probably a result of the social services absence and the consequent elimination of working agents due to starvation.

For the funds available per each pensioner, however, not surprisingly, it is the exact opposite. Whereas the fixed fee and pension taxes flow to the fund and population decreases simultaneously, fund deposits will increase, as well as the overall share of each pensioner.

When social services are available, and in an attempt to build a more realistic model, productivity decay has been taken into account, and the genetic traits have been set to follow a normal distribution, our quadruple indicators demonstrate considerable cyclic patterns as previously mentioned, the effects of the change in arbitrary parameters can be clearly observed in the amplitude and magnitude of population waves, not the mean or tendent value which they oscillate around, where both of the wave's amplitude and period length are in reverse relation with the retirement age.

The available amount of funds per retiree also generally decreases with the increase of mean maximum age, which makes sense since, in this case, more retired agents are alive at each time and each one's share drops. In this case, GDP per capita is also not affected by changes in the parameters, just like the Gini coefficient, but it is way less than in other circumstances.